\documentstyle[aps,epsfig,multicol,amstex,subfigure]{revtex}

\draft
\begin{document}
\title{Electron transport through quantum wires and point contacts}
\author{P. Havu, M. J. Puska, and
R. M. Nieminen} \address{Laboratory of Physics, Helsinki
University of Technology, P.O. Box 1100, FIN-02015 HUT, Finland }

\author{V. Havu} 

\address{ Institute of Mathematics, Helsinki University of Technology,
P.O. Box 1100, FIN-02015 HUT, Finland}

\maketitle
\begin{abstract}
We have studied quantum wires using the Green's function technique and
the density-functional theory, calculating the electronic structure
and the conductance. All the numerics are implemented using the
finite-element method with a high-order polynomial basis. For short
wires, i.e. quantum point contacts, the zero-bias conductance shows, as
a function of the gate voltage and at a finite temperature, a plateau
at around 0.7G$_0$.  (${\rm G}_0 = 2e^2/h$ is the quantum
conductance). The behavior, which is caused in our mean-field model by
spontaneous spin polarization in the constriction, is reminiscent of
the so-called 0.7-anomaly observed in experiments.  In our model the
temperature and the wire length affect the conductance-gate voltage
curves in the same way as in the measured data.
\end{abstract}


\begin{multicols}{2}
\section{Introduction}

Two-dimensional (2D) nanostructures can be fabricated at semiconductor
interfaces (such as GaAs/AlGaAs), using lithographic techniques and
gate electrodes \cite{Kane,Thomas,joachim}.  In these structures
conducting electrons form a quasi-2D electron gas at the
interface. Doping impurities are separated from the 2D electron gas
making the transport ballistic over the nanostructure. This enhances
the importance of quantum mechanical effects.

An interesting 2D nanostructure is the quantum wire (QW), which is a
laterally narrow electron pathway connecting two (infinite)
electrodes. Because of the ballistic electron transport the
conductance is quantized in units of the quantum $\rm{G}_0 =
2e^2/h$. The 2D electron states are quantized in the direction
perpendicular to the axis of the wire. Each occupied perpendicular
state gives rise to one conducting mode adding one quantum to the
conductance. In a typical measurement, the conductance is recorded as
a function of the gate voltage which lowers the potential in the wire
region, influencing its width or the average electron density. The
increase in the number of conducting modes results in a staircase as a
function of the gate voltage.  Recently, Kane et al. \cite{Kane} and
Reilly et al. \cite{Reilly} have measured conductances of QW's with
different geometries.  The quantization of the conductance is clearly
seen in the measurements.

A short quantum wire can be characterized as a quantum point contact
(QPC). QPC's exhibit the so-called 0.7-anomaly which is a small
plateau in conductance around 0.7-0.5~G$_0$ as a function of the gate
voltage \cite{Thomas}.  The plateau moves from 0.7~G$_0$ to 0.5~G$_0$
as the electron density in the quantum wire is lowered or the QPC is
in a strong magnetic field. The plateau becomes more pronounced
when temperature increases from the milli-Kelvin region to a few
Kelvin. At higher temperatures the anomaly disappears. The
length of the wire affects also the anomaly \cite{Reilly} so that in
long wires the anomaly is around 0.5~G$_0$ and in shorter ones around
0.7~G$_0$. The measurements show a short plateau also within the next
quantum step around 1.5~...~1.76~G$_0$.

There are several different explanations for the 0.7 anomaly. One of
them \cite{kondo,kondo2} is based on the Kondo model where an unpaired
electron is localized at the QPC. In the ground state this electron is
in the spin-singlet state with the scattering electrons from the
leads. The spin coupling results in a high density of states, a Kondo
resonance, at the Fermi level and thus in an enhanced conductance
around zero bias voltage. This zero-bias anomaly is similar to the
Kondo effect in single-electron transistors \cite{kondo_qd}. In the
case of the QPC the Kondo model seemingly explains the plateaus and
their behavior as a function of the temperature and the external
magnetic field \cite{kondo,kondo2}

Another explanation is provided by the semi-empirical model by Reilly
{\it et al.}  \cite{Reilly}. In this model the plateau is caused by
the local spin polarization of the electron gas at the QPC. For small
electron densities at the QPC the polarization vanishes, but when the
electron density is increased using the gate voltage a spin gap opens
and causes the polarization. The opening of the spin-gap depends on the QPC
configuration and temperature, and in different cases the conductance
plateaus appear at different values. A spin-polarization model has
also been presented by Berggren and Yakimenko \cite{berggren}. They
used the density-functional theory (DFT) to calculate the electron
density in a quantum wire as function of the gate voltage. The
conductance was calculated using the B\"uttiker model \cite{datta} with
parameters obtained by fitting the effective DFT potential in the
middle of the QW. Also Meir {\it et al.}  \cite{kondo2} and Hirose
{\it et al.} \cite{kondo} reported DFT calculations for a QW in order to
model the QPC resonance states needed in the Kondo model. Also recent
measurements utilising the resonant interaction between couplet QW's
show evidence of the formation of localized magnetic moment in a
constriction of the 2D electron gas \cite{ap103,prl04}.

In this work we investigate the spontaneous spin-polarization model
for QW's. Our aim is to make within this model accurate and realistic
calculations for the conductance, in order to study how far the model
can describe the experimental findings.  We use the DFT and the
Green's functions techniques (see, for example Xue {\it et al.}
\cite{Xue}) to calculate the electron density and the conductance. For
the electron spin densities the use of the Green's function
technique is computationally more demanding than solving for the wave
functions of a finite system. However, it has important
advantages. First, an infinite system without any artificial
periodicity can be treated. It is also possible to add the bias
voltage between the electrodes and calculate the charge density and the
current through the structure self-consistently. We solve for the
DFT-Green's function equations numerically by the finite-element
method using our recent implementation \cite{oma}.

We use the effective atomic units which are derived by setting the
fundamental constants $e = \hbar = m_e = 1$, and the material
constants $m^* = \epsilon = 1$. $m^*$ and $\epsilon$ are the relative
effective electron mass and the relative dielectric constant
respectively.  The effective atomic units are transformed to the usual
atomic and SI units with the relations
\begin{center}
\begin{tabular}{l l l l}
Length: & $1 \, a_0^*$ & $= 1 \frac{\epsilon}{m^*} a_0$ & $\approx
10.0307$ nm \\ Energy: & $1 \, {\rm Ha^*}$ &$= 1
\frac{m^*}{\epsilon^2} {\rm Ha}$ & $\approx 11.3079$ meV \\ Current: &
$1 \, {\rm a.u.}^*$&$ = 1 \frac{m^*}{\epsilon^2} {\rm a.u.} $ &
$\approx 2.7512$ $\mu$A.
\end{tabular}
\end{center}
Above, the numbers on the right-hand side are obtained by using the
parameters for GaAs, i.e. $m^*=0.067$ and $\epsilon=12.7$. They are
used in illustrating our results. Ha denotes the Hartree energy unit.

\section{Model and theoretical methods}

\begin{figure}[htb]
\begin{center}
\epsfig{file=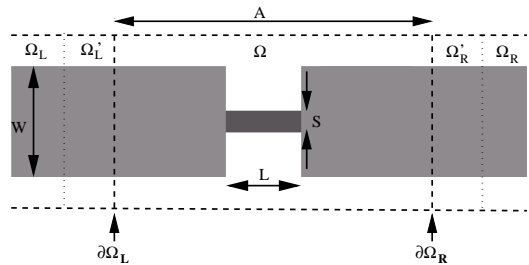,width=0.5\textwidth}
\end{center}
\caption{\label{kvanttikontakti} 2D quantum wire between two
  electrodes. The gray areas denote the rigid positive background
  charge. The electrodes continue to the infinity outside the
  calculation region. The uniform background charge density is varied
  in the dark-gray wire region in order to model the effects of the
  gate voltage.}
\end{figure}

Our strictly 2D model for a QW is shown in
Fig.~\ref{kvanttikontakti}. We use a simplified geometry in order to
see the effects due to different shapes of the wire.  The calculation
area, $\Omega$, consists of the QW and some parts of the electrodes.

The semi-infinite electrodes consist
of the positive background charge with the constant density of $0.2
(a_0^*)^{-2} \approx 2 \times 10 ^{11} e/\rm{cm}^2$ and the
neutralizing 2D electron gas with the density $\rho(r)$. 

At both sides of the electrodes and the wire we include some empty
vacuum, and the electron density is required to vanish at the outer
boundaries parallel to the wires. 

The electron density at the axis of the electrode is then typical for
GaAs/GaAlAs systems and the Fermi energy (the width of the occupied
energy band) is 0.63~Ha=7.1~meV.  The neutrality makes the calculation
of the electrostatic potential in the self-consistent DFT calculations
possible.  The QW is also modeled using a rigid uniform positive
background charge.  We control the positive charge density $\rho_j$ in
the wire region $\Omega_{QW}$ (the darker area in
Fig.~\ref{kvanttikontakti}) in order to mimic the effects of a gate
voltage.  The denser the background charge the deeper the effective
potential is, i.e.
\begin{equation}
V_g(r) = \int_{\Omega_{QW}} \frac{-\rho_j}{|r-r'|} dr', 
\end{equation}
where $r=\{r_x,r_y\}$ are the 2D coordinates. We calculate $V_g$ at
the midpoint of the wire and use this value as the gate voltage in our
illustrations below. The dependence of the gate voltage can be written
in the form
\begin{equation}
V_g = C_j \rho_j,
\end{equation}
where the coefficient $C_j$ depends on the shape of the wire. $C_j$ =
218mV$a_0^{*2}$, 234mV$a_0^{*2}$ and 248mV$a_0^{*2}$ for wires with
the width $W$ of 5$a_0^*$ and lengths $L$ of 6$a_0^*$, 7$a_0^*$ and
8$a_0^*$, respectively. In chapter III we denote these wires as D, E
and F.

We use the DFT within the local density-approximation to calculate the
electron density. In the self-consistent iterations the electron
density is first calculated using Green's functions. The electron
density is then used to calculate the effective potential $V_{eff}$ as
\begin{equation}
V_{eff}(r) = \int_{\Omega+\Omega_L'+\Omega_R'}
\frac{\rho(r')-\rho_+(r')}{|r-r'|} dr' + V_{xc}(r) + V_{bias}(r)
\end{equation}
Above, the first term on the right-hand side is the Coulomb potential,
computed as an integral and not from the Poisson equation because of
the 2D calculation space. The electron density in the
electrodes $\Omega_{L/R}$ is assumed to coincide with that in an
infinite uniform wire. Then we can include the outside regions into
the Coulomb integral within the buffer regions $\Omega_L'$ and
$\Omega_R'$ which are large enough.  For the exchange-correlation
potential, $V_{xc}$, we use the 2D-LDA functional by Attaccalite {\it
et al.} \cite{xc1,xc2}. The potential $V_{bias}$ is a linear ramp
taking care of the boundary conditions under a finite bias so that the
potential in the left electrode is shifted relative to that in the
right electrode by a given bias voltage $V_{SD}$. The effective
potential determines through the Green's functions a new electron
density and the procedure is repeated until convergence.

The retarded Green's function is obtained by solving the equation
\begin{equation}\label{greenR}
\big{(}\omega+\frac{1}{2}\nabla^2 - V_{eff}(r) \big{)}
G^r(r,r';\omega) = \delta(r-r'),
\end{equation}
where $\omega$ is the electron energy. At the boundaries $\partial
\Omega_{L/R}$ we use open boundary conditions given by the
Dirichlet-to-Neumann map on the boundary. This means that there are no
reflections. One can next calculate the so-called lesser Green's function,
$G^<(r,r';\omega)$. For zero bias, when the Fermi functions in
the left and right electrodes are identical, i.e. $f_L(\omega) =
f_R(\omega)$, $G^<$ is obtained as
\begin{equation}\label{ele1}
G^<(r,r';\omega) = 2f_{L/R}(\omega) G^r(r,r';\omega).
\end{equation}
In the non-equilibrium situation when $f_L(\omega) \neq f_R(\omega)$ we have
\begin{equation}
\begin{aligned}
\label{ele2}
 G^<&(r,r';\omega) = \\ & -i f_R(\omega) \int_{\partial \Omega_R}
\int_{\partial \Omega_R} G^r(r,r_R;\omega) \Gamma_R(r_R,r'_R;\omega)
\\ & \quad \quad \quad \quad \quad \quad \quad \times
G^a(r'_R,r';\omega) \, dr_R \, dr'_R \\ &-if_L(\omega) \int_{\partial
\Omega_L}\int_{\partial \Omega_L} G^r(r,r_L;\omega)
\Gamma_L(r_L,r'_L;\omega) \\ & \quad \quad \quad \quad \quad \quad
\quad \times G^a(r'_L,r';\omega) \, dr_L \, dr'_L.
\end{aligned}
\end{equation}
Above, $\Gamma_{L/R}$ are functions connecting the calculation area to
the outside electrodes. They have nonzero values only on the
boundaries $\partial \Omega_{L/R}$. The imaginary part of $G^<$ is
related to the density of states, and the electron density is
calculated by integrating over the energy
\begin{equation}\label{ele_integraali}
\rho(r) = \frac{-1}{2\pi} \int_{-\infty}^{\infty}
\rm{Im}(G^<(r,r;\omega)) d\omega.
\end{equation}
The local density of states (LDOS) in the QW region is calculated as
\begin{equation}\label{ldos_eq}
g_{QW}(\omega) = \frac{-1}{2\pi} \int_{\Omega_{QW}}
\rm{Im}(G^<(r,r;\omega)) dr.
\end{equation}
LDOS is a continuous function in our calculations and thus the
finite-size effects are small.

The electron tunneling probability $T(\omega)$ is calculated as a
function of the electron energy using the Green's functions as
\begin{equation}\label{tunneling_probability}
\begin{aligned}
T(\omega) =& \int_{\partial \Omega_L} \int_{\partial \Omega_L}
\int_{\partial \Omega_R}\int_{\partial \Omega_R} 
\Gamma_L(r_L,r'_L;\omega) 
 G^r(r'_L,r_R;\omega ) \\
\times& \Gamma_R(r_R,r'_R;\omega)
G^a(r'_R,r_L;\omega)
\, dr_L \, dr'_L \, dr_R \,dr'_R,
\end{aligned}
\end{equation}
Where $G^a$ is Green's advanced function. Thus one needs the Green's
function values only at the boundaries $\partial \Omega_{L/R}$. For a
finite bias the electric current is calculated as
\begin{equation}
I = \frac{1}{\pi} \int_{-\infty}^{\infty} T(\omega) \left( f_L(\omega) -
f_R(\omega) \right) d\omega,
\end{equation}
where the Fermi functions $f_L$ and $f_R$ are shifted with respect to
each other by the bias voltage $V_{SD}$. In the zero bias limit
$f_L(\omega) = f_R(\omega) = f(\omega)$ and one obtains the linear-response
conductance
\begin{equation}\label{G_temp}
G =  \frac{1}{\pi} \int_{-\infty}^{\infty} T(\omega) \frac{d
  f(\omega)}{d\omega} d\omega.
\end{equation}
At zero temperature the conductance is simply $T(\omega_f)$, where
$\omega_f$ is the Fermi energy. At a finite temperature also electron
states with energies near the Fermi level contribute to the
conductance, as the derivate of the Fermi function $f(\omega)$ differs
from the delta function. A finite temperature influences the solution
also through the electron density (cf. Eqs
(\ref{ele1}) and (\ref{ele2})). Below we show also differential
conductances corresponding to given bias voltages. At zero
temperature they can be computed from the approximation
\begin{equation}
\frac{dI}{dV_{sd}} \approx \frac{I(V_{sd}+\delta V) - I(V_{sd})}{\delta V}
\end{equation}
where $\delta V$ is a small increment in the bias voltage.

We have numerically implemented the non-equilibrium DFT-based scheme
using the finite-element method as explained earlier \cite{oma}. The
size of the matrices to be inverted is determined by the number of the
finite-element basis functions needed in the calculations. We have
implemented 2D high-order polynomial bases \cite{p} up to the fourth
order in order to reduce the basis size. In a typical calculation the
number of basis functions ranges from 2800 for to high-order
polynomials to 5500 for low-order polynomials.

\section{Results and discussion}

\subsection{Spin polarization in the QW}

\begin{figure}[htb]
\begin{center}
\epsfig{file=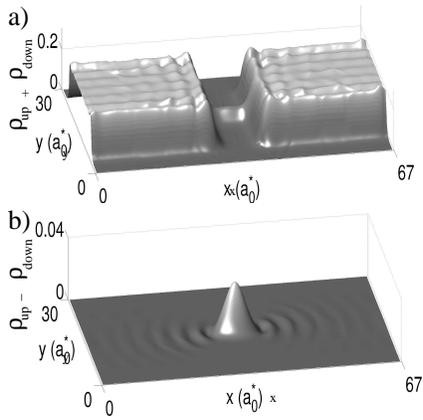,width=0.40\textwidth}
\end{center}
  \caption{ (a) Total electron density and (b) difference between the
  spin-up and spin down electron densities at the zero temperature for
  the system with the dimensions S=5~a$_0^*$, L=7~a$_0^*$,
  W=20~a$_0^*$, A=47~a$_0^*$ (see Fig.~\ref{kvanttikontakti}) and for
  the gate voltage $V_g$= 12.4~mV}
  \label{spin}
\end{figure}

In this model, there are electronic resonance states in the QW, which
are reflected as peaks in the LDOS (see Fig.~\ref{LDOS} below). The
peaks are broader in short and wide QW's than in long and thin ones,
because short QW's are more strongly connected to the electrodes. If
the resonance peaks are narrow enough, a spontaneous spin polarization
may occur with a spin gap opening between the spin-up and spin-down
states below and above the Fermi level, respectively.  The
polarization occurs in a limited range of gate voltages. At small gate
voltages there is no polarization and it disappears again at large
voltages. This is in accordance with the density-dependent opening of
the spin gap speculated by Reilly {\it et al.} \cite{Reilly}.  A
solution with spin polarization in the QW is shown in Fig.~\ref{spin}
which gives the total electron density and the difference between the
spin-up and spin-down densities. The electron wave functions have in
this case no nodes in the direction perpendicular to the QW, so that
there is only a single conducting mode.

Berggren and Yakimenko \cite{berggren} have studied the spin
polarization in QW's.  In their calculations the electrodes are
modeled by two large quantum dots which are connected by the QW. The
potential and the electron density inside the QW are controlled by a
gate voltage. The electron density is calculated from the wave
functions using the DFT within the LDA.  They obtain a spin
polarization which closely resembles our results. The spread of the
polarization in the electrodes is somewhat larger than in our cases
which show only a small oscillating polarization outside the QW.

Berggren and Yakimenko do not report on the details of the resonance
peaks in the DOS. The resonance state parameters are crucial in the
Kondo model. Therefore Meir {\it et al.} \cite{kondo2} have used the
DFT to model the QPC. In their calculations the semi-infinite
electrodes are modeled by wires with a parabolic confinement potential
perpendicular to the wire axis, and a raise in the external potential
forms the QPC as a saddle-point. Due to the different potential
construction the resonance peaks obtained by Meir {\it et al.} are
remarkably wider than those in our calculations.  A similar QPC
construction is used also in the calculations by Hirose {\it et al.}
\cite{kolme}. The polarizations they obtain are similar to our results
for long QW's.


\subsection{Conductance as function of the width of the wire}

\begin{figure}[htb]
\begin{center}
\epsfig{file=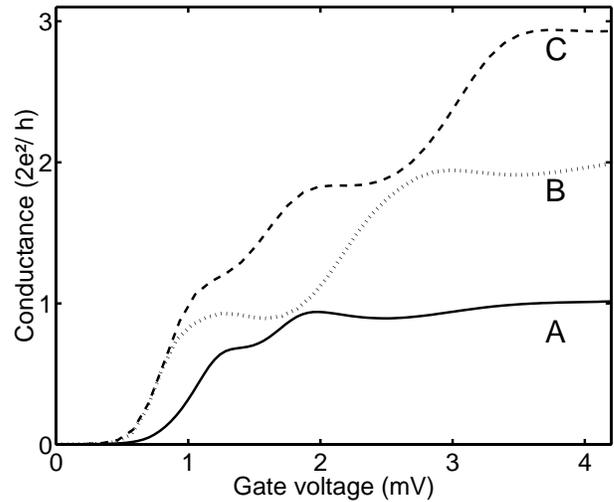,width=0.45\textwidth}
\end{center}
\caption{\label{monta} Conductance at zero temperature as a function
of the gate voltage for three wires with the length L=7~a$_0^*$ and
widths S=5~a$_0^*$ (A), 6~a$_0$ (B) and 10~a$_0^*$ (C).  The width
of the electrodes is W=20~a$_0^*$, and the length of the computational
area is A=47~a$_0^*$ (see Fig.~\ref{kvanttikontakti}.).}
\end{figure}

The conductances of three QW's A, B and C with different widths
are shown in Fig.~\ref{monta} as a function of the gate voltage at
zero temperature. The QW's have the same length L=7~a$_0^*$ but
for A, B, and C the widths S=5~a$_0^*$, 6~a$_0^*$ and 10~a$_0^*$,
respectively. The staircase quantizations of the conductances are
clearly seen. For the widest wire C the perpendicular states are
denser in energy than for the thinner wires A and B, so that the
conductance steps are shortest for C. The first plateau for C is also
disturbed by the states decaying from the electrodes to the QW. Wires
A and B are connected more weakly to the electrodes and for them the
first plateau is clearer. A weak connection causes also the
spin-polarization in wire A, seen as a short plateau clearly below
1~G$_0$. Wire C is not polarized and in wire B the polarization is not
strong enough to cause plateaus deviating from integer of G$_0$ values.

The gate voltage drives the electrostatic potential quite uniformly
within the QW. In experiments, side gates are often used to control
the potential \cite{Kane}. These gates not only increase the potential
in the QW, but they also make it narrower. The result is a conductance
staircase as a function of the gate voltage. The first measured
plateaus corresponding to 1G$_0$ are pronounced \cite{Kane} in
contrast to our shoulder-like first plateau for wire C.  Below we are
concerned mainly with the conductance plateaus below 1G$_0$ and
therefore we discuss wires with the width S=5~a$_0^*$ (wire A in
Fig.~\ref{monta}).

\subsection{Conductance as function of the length of the wire}

\begin{figure}[htb]
\begin{center}
\epsfig{file=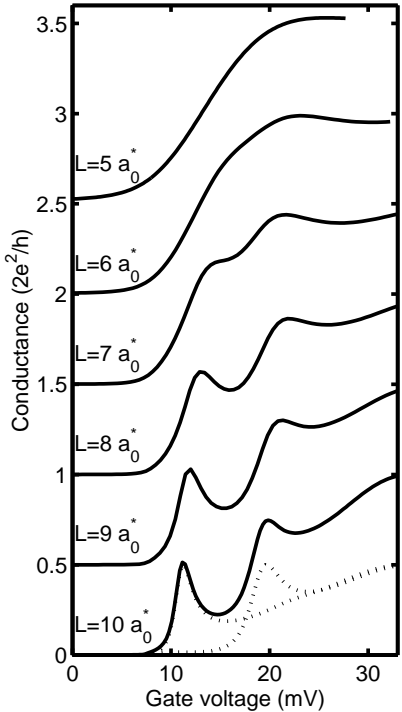,width=0.3\textwidth}
\end{center}
\caption{\label{cont_lenght} Conductance as a function of the gate
  voltage for QW's with the width S=5~a$_0$ and with different lengths
  L.  The width of the electrodes is W=20~a$_0^*$, and the length of the
  computational area is A=47~a$_0^*$ (see also
  Fig.~\ref{kvanttikontakti}). The successive curves have been shifted
  by 0.5~G$_0$. The conductance of the wire with L=10~a$_0$ is
  decomposed into spin-up and spin-down contributions (dotted lines).}
\end{figure}

The conductances of the wires with the width S=5~a$_0^*$ and with
different lengths L are shown in Fig.~\ref{cont_lenght} as a function
of the gate voltage and at zero temperature.  The figure shows clearly
the effect of the electrode-wire connection. The long wires have clear
peaks due to the resonances. The heights of the peaks are 0.5~G$_0$,
meaning that only a single electron polarized mode contributes to
them.  The wires with the lengths L=6...8~a$_0^*$ exhibit resonances
which are just narrow enough for the spin polarization to appear.  The
length dependence of the conductance among these three wires is in
qualitative agreement with the recent measurements for QW's by Reilly
{\it et al.} \cite{Reilly}, although the wires in the experiments are
clearly longer than in our calculations.

Berggren and Yakimenko \cite{berggren} have also calculated the
conductance of their QW system using a rough scheme. They fit
parabolic curves to the effective potential in the middle of the
QW. The fitting parameters are then used in the B\"uttiker model for
the conductance.  The results have the same overall features as ours
for the longest wires in Fig.~\ref{cont_lenght}, but the changes in
the polarization and in the conductance are less abrupt in our
calculations. Berggren and Yakimenko argue that the rapid changes are
due to the finite-size effects.  Therefore our results describing
infinite systems seem to support their statement.  The
conductance-gate voltage curves by Berggren and Yakimenko show two
plateaus, one plateau when the polarization appears and another one at
$\sim$0.75~G$_0$, just before the polarization disappears.  This
result resembles our curves for the long wire with
L=10~$a_0^*$. However, there is a notable difference that in our
calculations for the long wires the polarization at low gate voltages
is first nearly perfect whereas that by Berggren and Yakimenko is
first absent and appears then as a partial polarization.  In our
conductance curves only one plateau is seen when the polarization
reaches its maximal value.

\subsection{Temperature dependence of the conductance curves}

\begin{figure}
  \centering
  \mbox{\subfigure{\epsfig{figure=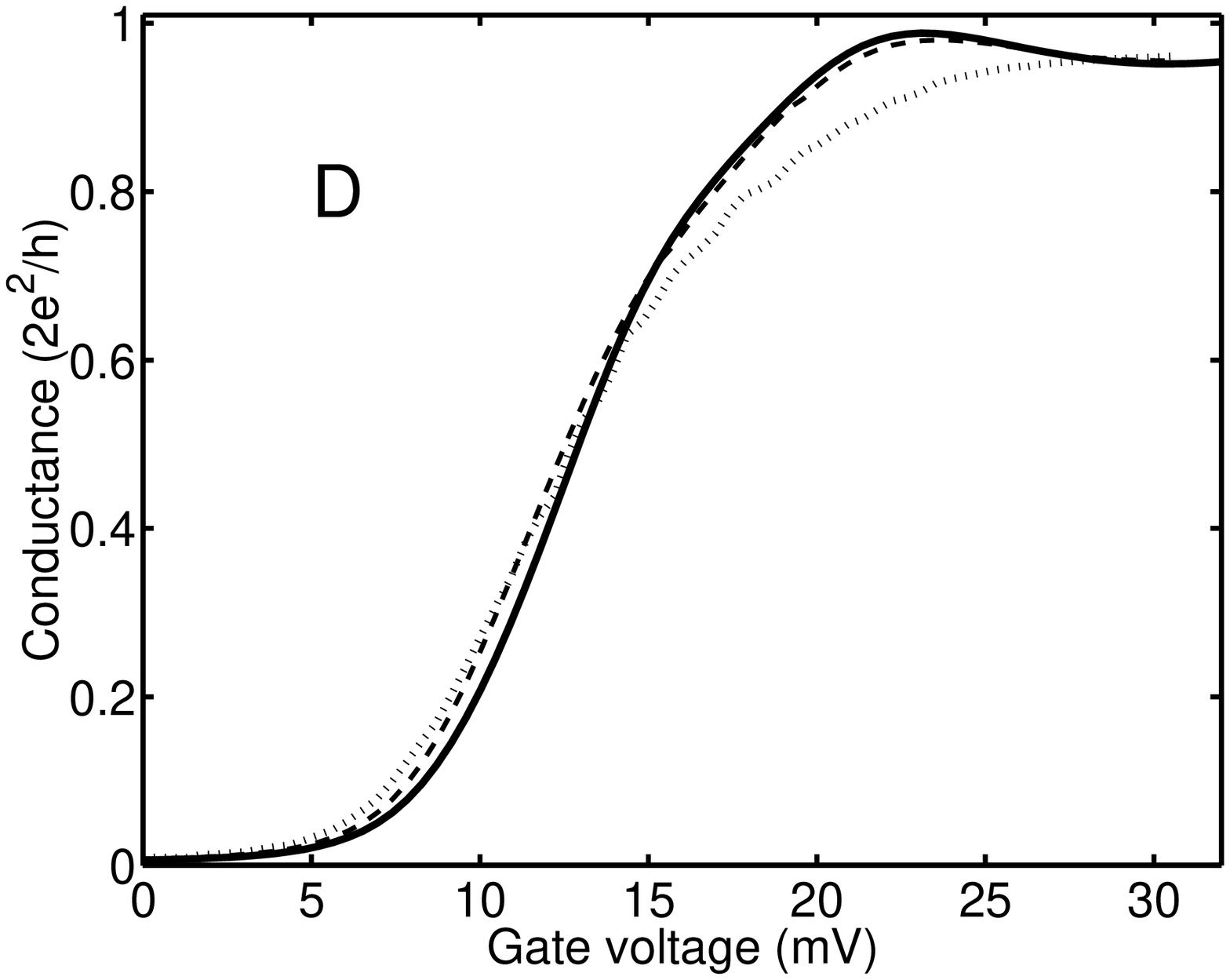, width=0.4\textwidth}}}\\
  \mbox{\subfigure{\epsfig{figure=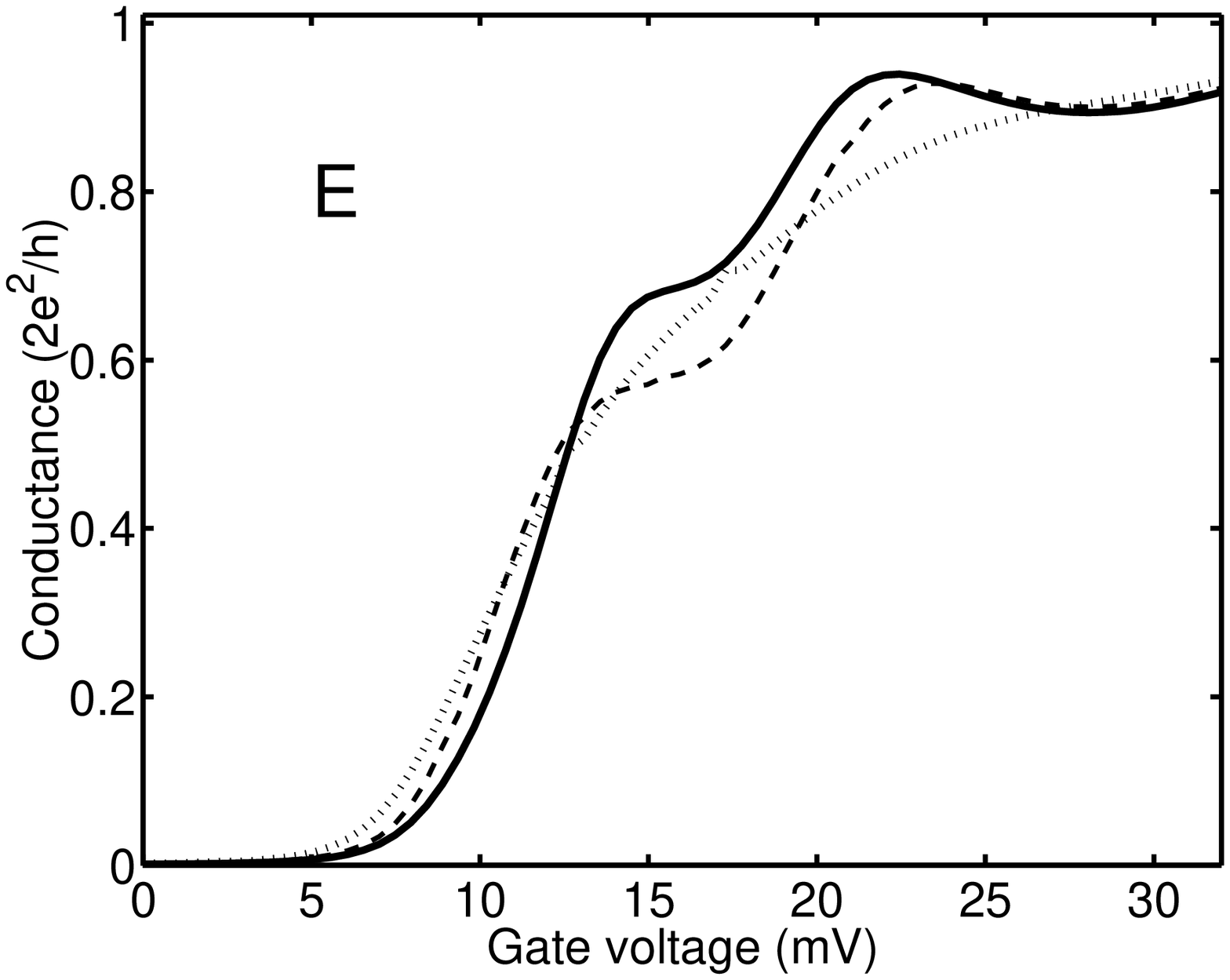, width=0.4\textwidth}}}\\
  \mbox{\subfigure{\epsfig{figure=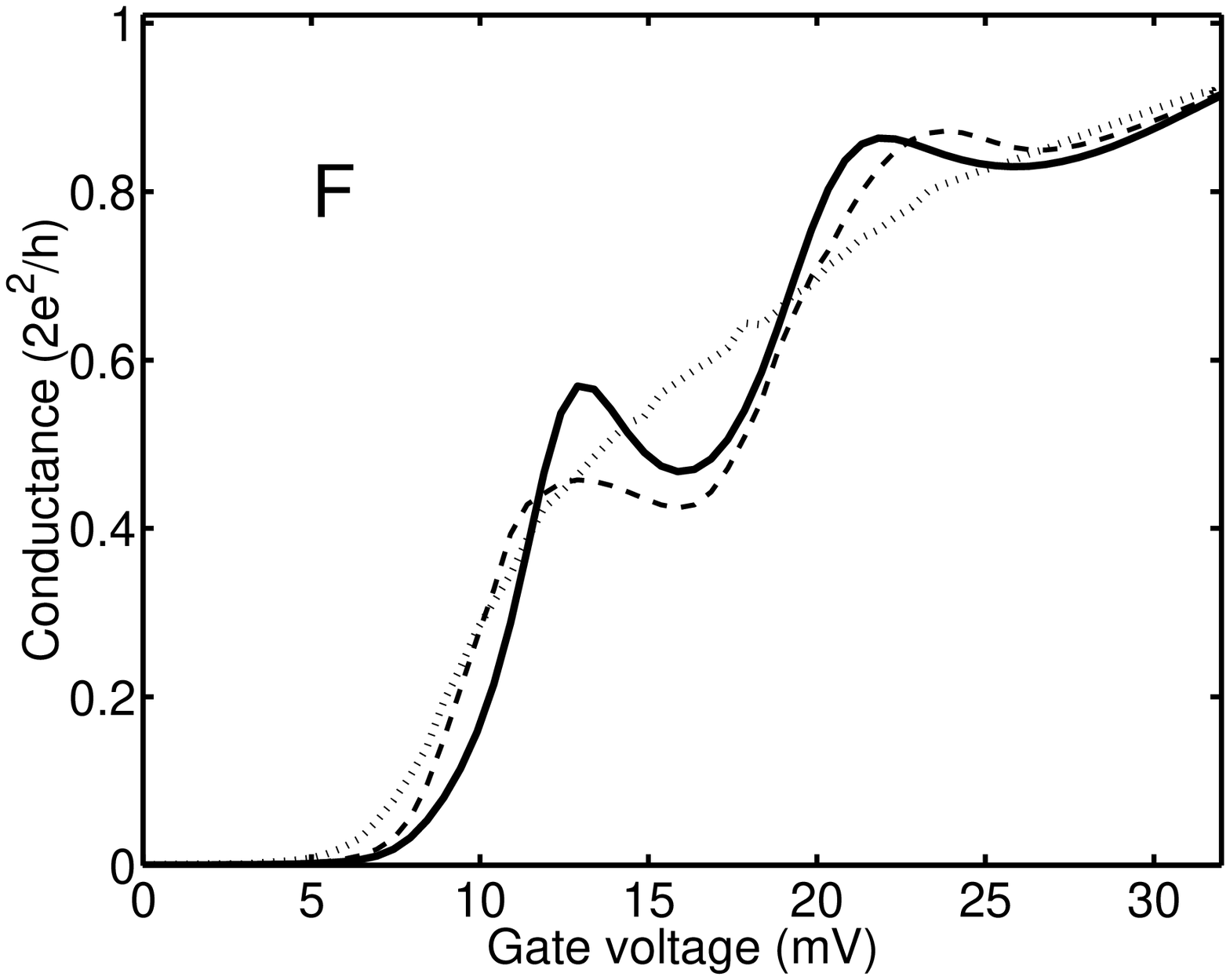,width=0.4\textwidth}}}
  \caption{Conductance as a function of the gate voltage for QW's with
  the width S=5~a$_0^*$ and lengths L=6~a$_0^*$ (D), 7~a$_0^*$ (E)
  and 8~a$_0^*$ (F) at temperatures of 0~K (solid curve), 2~K
  (dashed curve) and 4~K (dotted curve).}
  \label{gate_temp}
\end{figure}

The effect of the temperature on the conductance behavior of three
QPC-like wires D, E and F is shown in Fig~\ref{gate_temp}. At zero
temperature, wire D (L=6~a$_0^*$) shows no plateau, whereas wire E
(L=7~a$_0^*$) has a plateau at $\sim$0.7~G$_0^*$ and wire F
(L=8~a$_0^*$) at $\sim$0.5~G$_0$. When temperature increases the
plateaus below $1\rm{G}_0$ in wires E and F shift down wards and become
smoother. Wire D shows a weak temperature dependence so that the slope
at $\sim$0.7G$_0$ decreases.

Reilly {\it et al.} \cite{nelja} have presented a phenomenological
model for the appearance and the location of the conductance
plateaus. They report on the measurements of three QPC's and explain
their conductance behaviors. Two of the QPC's have plateaus at
$\sim$0.5~G$_0$ and $\sim$0.7~G$_0$ and the third one has a very weak
plateau at $\sim$0.6~G$_0$. The QPC's, D, E and F in
Fig.~\ref{gate_temp} show similar gate-voltage and temperature
dependences.

In the model of Reilly {\it et al.} the relevant parameter is the
ratio between the spin gap width and the thermal energy $k_BT$.
According to our calculations the temperature broadening in the Fermi
functions is small when compared to the resonance peak widths arising
from the interactions with the electrodes. Because of this, $k_B T$
has to be replaced by the resonance peak width as the relevant
parameter. Then the model can be used to explain the
conductance-gate voltage curves also at zero temperature.

Reilly {\it et al.} \cite{nelja} have given three scenarios based on their
model. In their scenario I the spin polarization increases rapidly
with the increase of the electron density at the QPC. Then a plateau
appears at $\sim$0.5~G$_0$. This corresponds to our wire F, in which the
resonance peaks are narrowest.
In the scenario II by Reilly {\it et al.} the temperature broadenings
(the widths of the resonance peaks in our generalization) are
comparable to the spin gap. This corresponds to our wire E. When the
spin gap begins to open at a certain gate voltage also the spin-down
resonance states are partially populated. This situation remains when
the density increases and results in the conductance plateau at
$\sim$0.7~G$_0$.
In scenario III by Reilly {\it et al.} the spin splitting is weak
in comparison to the width of the resonance peaks. This corresponds to our
wire D for which no plateau is seen at low temperatures.

\begin{figure}[htb]
\begin{center}
\epsfig{file=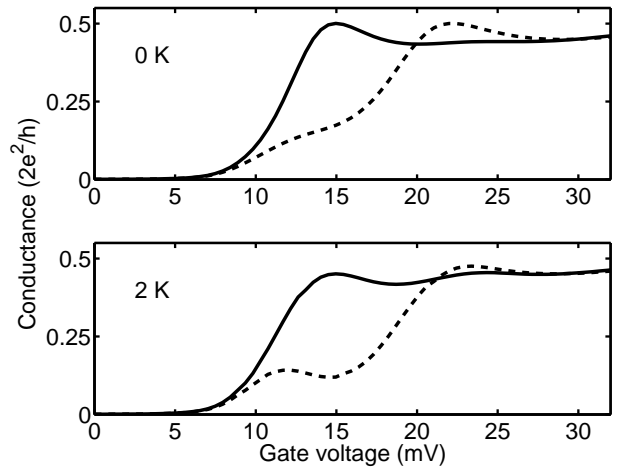,width=0.45\textwidth}
\end{center}
\caption{\label{gate6_diff} Conductances due to spin-up (dashed
  curves) and spin-down (solid-curves) electrons as a function of the
  gate voltage and at temperatures of 0~K and 2~K. The results are
  for wire E (Fig.~5) with the length L=6.5~$a_0^*$.}
\end{figure}

\begin{figure}[htb]
\begin{center}
\epsfig{file=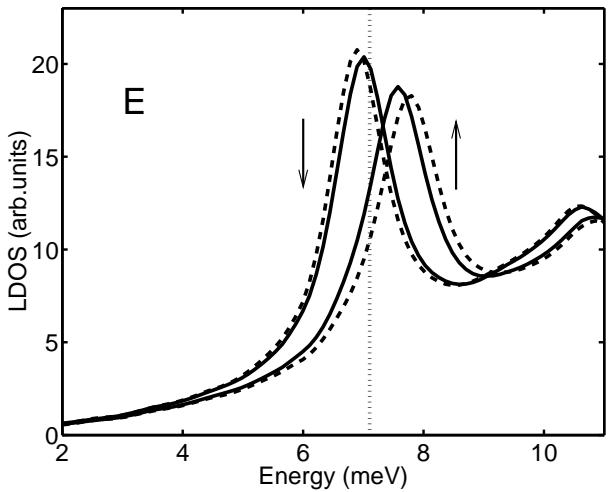,width=0.45\textwidth}
\end{center}
\caption{\label{LDOS} LDOS corresponding to the QW region
  (Eq.~\ref{ldos_eq}) wire E with the width S=5~$a_0^*$ and length
  L=7~$a_0^*$. The decomposition to the spin-up and spin-down states at
  the gate voltage of 14~mV and at the temperatures of 0~K
  (solid-curve) and 2~K (dashed-curve) are given. The dotted line
  denotes the Fermi-level.}
\end{figure}

The reason for the different temperature behaviors of the QW's in our
model can be seen from the curves in which the conductances caused by
spin-up and spin-down electrons are separated. For wire E this is
shown in Fig.~\ref{gate6_diff}. As temperature increases from 0~K
to 2~K spin polarization increases for gate voltages around the middle of
the plateau below 1G$_0$. At the same time, the resonance peaks also become
wider because more states contribute to the conductance (see
Eq.~(\ref{G_temp})).  The same behavior is seen also for wires D and F.
The reason for the increase in the polarization is seen in the LDOS
for wire E in Fig.\ref{LDOS}. 
When the temperature rises the electron density increases in the QW
due to the resonances near the Fermi level. Then the decrease in the
exchange-correlation energy opens the spin gap, as can be seen in
Fig.~\ref{LDOS}, and the polarization increases.
The effect is clearer for wire E than for wires D and F. In wire D the
electron density does not increase as fast as in wire E because the
resonance peaks are wider. Wire F has a strong polarization already at
zero temperature and therefore it cannot show an increase as large as
wire E.

Meir {\it et al.} \cite{kondo2} have modeled a QW's using the Kondo
model \cite{kondo2} as explained in the Introduction. In their model
the Anderson Hamiltonian with parameters controlling the properties of
the QPC is used. In their example the conductance plateau is smooth at
low temperatures and located around $\sim$0.9~G$_0$. As the
temperature increases the plateau becomes wider and moves down to
$\sim$0.7~G$_0$. The behavior is similar to that observed here for
wire E.

\subsection{Effect of the bias voltage}

\begin{figure}[htb]
\begin{center}
\epsfig{file=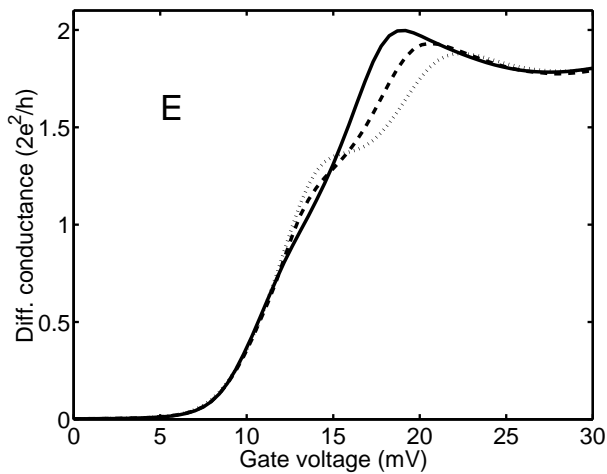,width=0.45\textwidth}
\end{center}
\caption{\label{diff_gate} Differential conductance of wire E with the
  width S=5 a$_0^*$ and length L=7$a_0^*$ as a function of the gate
  voltage and at zero temperature. Results for the bias voltages of
  0~mV (dotted curve), 0.23~mV (dashed curve) and 0.46~mV (solid
  curve) are given.}
\end{figure}

The differential conductance calculated using Eq. (11) is shown in
Fig.~\ref{diff_gate} as a function of the gate voltage. The increase
in the applied bias voltage $V_{SD}$ mainly increases the conductance
but at the same time it also curtails the conductance plateaus.  This
is not exactly in agreement with the measurements which show that the
conductance plateau below 1~$G_0$ rises with the increasing bias
voltage \cite{mit1,kondo}. According to our calculations the raise of
$V_{sd}$ causes also the decrease in the spin polarization in the QW.

The measured differential conductance suppresses rapidly as a function
of the source-drain voltage. This is called as the zero-bias
anomaly. Our mean-field model cannot produce this kind of behavior
whereas the Kondo model \cite{kondo2} gives an explanation for the
zero-bias anomaly. In the Kondo model an unpaired electron couples to
the electrons of both leads, resulting in an increased LDOS (Kondo
resonance) at the Fermi levels of the electrodes and in an enhanced
conductance. When the separation between the Fermi levels increases
with bias voltage, electrons can no longer resonantly tunnel through
the QPC which leading to the suppression in the conductance.


\section{Conclusions}

We have used the density-functional theory and the Green's function
formalism to model electronic transport through ballistic 2D quantum
wires.  Our results show the appearance of a spontaneous spin
polarization in the wire. Under proper conditions, the spin
polarization causes a plateau at around 0.7~G$_0$ in the conductance
as a function of the gate voltage. The calculations explain
measurements by Kane {\it et al.}  \cite{Kane} for the movement of the
conductance plateau as function of the wire length. Also the
temperature dependence is qualitatively similar to the experimental
findings. As a mean-field theory our model can not explain the zero
bias anomaly in the conductance. However, the parameters of the
resonance state causing the spin-polarization can be used in the Kondo
model \cite{kondo}.

\acknowledgments

We acknowledge the generous computer resources from the Center for
Scientific Computing, Espoo, Finland.  This research has been
supported by the Academy of Finland through its Centers of Excellence
Program (2000-2005). P.H. acknowledges financial support by Vilho,
Yrj\"o and Kalle V\"ais\"al\"an foundation.



\end{multicols}
\end{document}